# Access to Taxicabs for Unbanked Households: An Exploratory Analysis in New York City

Juan Francisco Saldarriaga
Columbia University
New York, NY, USA
jfs2118@columbia.edu

David A. King
Arizona State University
Phoenix, AZ, USA
davidandrewking@gmail.com

**ABSTRACT**

Taxicabs are a critical aspect of the public transit system in New York City. The yellow cabs that are ubiquitous in Manhattan are as iconic as the city's subway system, and in recent years green taxicabs were introduced by the city to improve taxi service in areas outside of the central business districts and airports. Approximately 500,000 taxi trips are taken daily, carrying about 800,000 passengers, and not including other livery firms such as Uber, Lyft or Carmel. Since 2008 yellow taxis have been able to process fare payments with credit cards, and credits cards are a growing share of total fare payments. However, the use of credit cards to pay for taxi fares varies widely across neighborhoods, and there are strong correlations between cash payments for taxi fares and the presence of unbanked or underbanked populations. These issues are of concern for policymakers as approximately ten percent of households in the city are unbanked, and in some neighborhoods the share of unbanked households is over 50 percent. In this paper we use multiple datasets to explore taxicab fare payments by neighborhood and examine how access to taxicab services is associated with use of conventional banking services. There is a clear spatial dimension to the propensity of riders to pay cash, and we find that both immigrant status and being 'unbanked' are strong predictors of cash transactions for taxicabs. These results have implications for local regulations of the for-hire vehicle industry, particularly in the context of the rapid growth of services that require credit cards. Without some type of cash-based payment option taxi services will isolate certain neighborhoods. At the very least, existing and new providers of transit services must consider access to mainstream financial products as part of their equity analyses.

## 1.INTRODUCTION

Taxicabs represent an important transit service in urban areas, and the industry undergoing rapid change. In recent years new technologies and private firms have shown substantial interest in growing the taxi industry from niche markets that complement transit systems to full-fledged alternatives auto ownership. Much of the current interest in taxicabs is connected with making taxicabs easier to use through smart phone based e-hail applications and credit card payments. While these innovations have no doubt made taxi services—both conventionally regulated taxicabs and upstart tech-oriented taxi services—easier to use for many travelers, these same innovations may make it harder for certain people to access them. In many U.S. cities large portions of low-income households do not have access to mainstream bank accounts or credit cards. These un- or under-banked households are effectively excluded from new services, fare discounts for transit passes, and other transportation services that require access to credit cards.

The literature on the underbanked rarely mentions transportation. Within the transportation literature, ability to pay is generally considered a function of income or wages. In the case of the underbanked, however, ability to pay must also include the fare payment media. There have been some studies where scholars have examined how the adoption of smart cards for transit fares may be affected by income, immigrant status and other factors (Yoh, Iseki et al. 2006). Within transit payments, low income riders tend to not take advantage of volume discounts or unlimited fares, which is likely caused by their precarious financial straits. But all of these examinations assume that users are at least *able* to access transit services or other transportation facilities. In the case of private taxi and transit services a lack of a formal bank account and credit card (or branded pre-paid debit card) prohibit the use of these services, at least in the United States.

Within the context of the un- and underbanked, there are many reasons why they may stay out of the formal banking system. First, they might not have employment with regular paychecks. If someone works odd jobs for cash they may not need an account for savings. Second, they may have a regular job with a steady paycheck, but the fees charged for bank accounts with a debit card are too high for their income or they receive most of their wages in cash tips. These people are likely to use check cashing stores, and paying check cashing fees may actually be cheaper than using an ATM throughout the week. Third, immigrants—both legally in the country and illegally—are less likely to have formal bank accounts than native born people. The reasons for this are not well understood beyond the obvious factors associated with immigrant status. Together, low income and immigrant status are associated with most of the un-and underbanked populations.

In 2010 New York City estimated that over 10 percent of the adult population was without a bank account (Sarlin and Miller 2010). The share of unbanked varies widely across the city, however, with nearly 30 percent of the population of the Bronx—the poorest borough—unbanked, while wealthier Staten Island has less than two percent unbanked (Empowerment). Moreover, nearly half of all unbanked live in just ten neighborhoods, all clustered in the poorest parts of the city, and happen to be places that have traditionally been underserved by taxicabs.

In the past few years the city has launched a series of programs with the cooperation of financial institutions to increase access to mainstream services (New York City Department of Consumer Affairs 2008). These programs have had modest success for encouraging saving, even among very low income people, and modest success moving people into mainstream accounts (New York City Department of Consumer Affairs 2013). Under the current Mayor de Blasio administration the city has created a municipal identification card that does not require citizenship to acquire. This new ID card is hoped to assist at least some of the unbanked population to open new accounts. Yet for all of the city's efforts, the evidence is mixed on the overall effectiveness of such "lifeline" services for promoting shifts into formal banking (Doyle, Lopez et al. 1998).

Overall, the concern presented here is that a particular aspect of poverty—whether or not a household has access to a formal bank account and thus potential access to credit cards—is increasingly important for available transportation choices. While some travelers pay cash due to privacy concerns, most who pay cash do not have an option. These can lead to higher fares and worse service, which may further isolate vulnerable communities. To the best of our knowledge access to bank accounts and credit cards has only marginally been addressed in the literature. This exploratory analysis uses taxi data from New York City to identify spatial factors associated with the likelihood of being unbanked and cash fares for taxi trips.

## 2. UN-AND UNDER BANKED IN NEW YORK CITY

This research uses specific definitions of un- and under-banked households. Unbanked households are those that do not have a checking or savings account in a formal bank. Underbanked may have access to mainstream banking facilities but due to language, physical proximity to bank branches or other reasons may not use bank accounts and products by choice, such as with migrant workers, or circumstance, such as an elderly household who physically can't reach the branch.

Many households go between banked and unbanked depending on their circumstance. In general there are a few factors strongly associated with being unbanked. The largest predictor of becoming unbanked is a steep decline in household income, followed by race and ethnicity factors, marital status and housing characteristics (Rhine and Greene 2013). Most of these factors are found concentrated in particular neighborhoods, which suggests that households on the edge of poverty in certain communities will move in and out of the banking system as they can afford to.

The extent of underbanking has recently been recognized around the world, but both the diagnoses and remedies depend greatly on local context. A few generalized statements about underbanked households can be made. They are more likely to be poor, both by income and wealth, than households with bank accounts. In the literature, the primary concerns about the unbanked are usually about the high costs of being poor, especially as this relates to the cost of money. Check cashing services can be more expensive than a savings account, for instance, as is getting a money order to pay all bills. Recently there has been some interest in pre-paid debit cards as a financial tool for low income families, but there are few examples of how these may work for transportation in the United States.

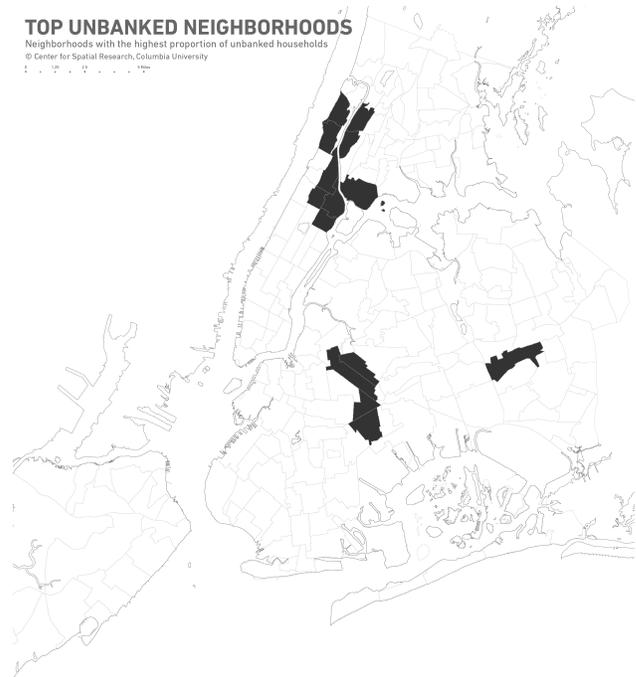

In New York City, studies show that physical proximity to conventional bank branches is unrelated to likelihood of being unbanked. Throughout the city bank branches are ubiquitous, though new bank branches are viewed as a sign of gentrification. While not the focus of this paper, the neighborhoods with high levels of unbanked households have mixed experience with gentrification, however it is defined.

Figure 1 shows the neighborhoods with the highest share of unbanked households shaded in grey along with subway stations throughout the city. There are approximately three clusters of unbanked communities: northern Manhattan and the south Bronx; east-central Brooklyn; and Jamaica, Queens. The neighborhoods in Manhattan and the Bronx are the poorest neighborhoods in these boroughs, but also have fairly good transit access by subway. Table 1 shows the actually share of unbanked households. Together, these ten neighborhoods represent about 450,000 unbanked people, or half of all unbanked in the city.

| Table 1: Largest Share of Unbanked Households by Neighborhood | | |
|---|---|---|
| **Name** | Borough | Unbanked |
| **Mott Haven/Melrose** | Bronx | 56% |
| **Morris Heights/University Heights** | Bronx | 53% |
| **Highbridge/Concourse** | Bronx | 51% |
| **Ocean Hill/Brownsville** | Brooklyn | 47% |
| **Bushwick** | Brooklyn | 47% |
| **Washington Heights/Inwood** | Manhattan | 46% |
| **West Harlem** | Manhattan | 38% |
| **East Harlem** | Manhattan | 37% |
| **Central Harlem** | Manhattan | 36% |
| **Jamaica** | Queens | 24% |

Source: Sarlin and Miller 2010

## 3. THE HIGH COST OF BEING POOR

Poverty is a major urban policy concern. For much of the post-war period in the United States poverty was largely an inner city phenomenon within metropolitan areas. One reason for concentrated poverty in the urban core was the availability of public transportation (Glaeser, Kahn et al. 2008). While poor, these households at least had access to transit networks that may allow for economic mobility, though our knowledge of how transportation affects poverty is limited (Sanchez, Shen et al. 2004, Sanchez 2008). In recent years, in part due to the Great Recession, poverty has suburbanized (Kneebone 2010). This has led to new concerns about the role of transit in suburban locations to prevent economic isolation for those who cannot afford to drive. But the costs associated with poverty are not limited to transportation options.

Poor households face a number of ways that reinforce how expensive it is to be poor. Inner-city neighborhoods pay higher retail prices (Talukdar 2008), for instance, or pay higher transit fares because they can't take advantage of discounts. WNYC, a news radio station in New York City, used data from the Metropolitan Transportation Authority to demonstrate where riders purchases 7-day transit passes for $30 or unlimited transit passes for $112 per month (SChuerman 2015). The MTA data shows that the 7-day passes are used more frequently than the unlimited passes, at 2.3 rides per day versus 1.9. This means that the average fare paid is somewhat less for the typical 7-day pass holder, the higher usage means that these riders would receive substantial discounts simply by switching to a monthly unlimited pass. It is not known precisely why transit riders purchase 7-day passes when unlimited passes would ultimately save them money, but the most likely explanation is that the travelers simply do not have $112 to commit to transit trips at the beginning of each month. What these riders can do is buy a shorter pass when they are able, and if not they don't travel or find other alternatives. This is a subtle example of how costs of living increase as income drops.

## 4. THE INFORMAL TRANSIT MARKET IN NEW YORK

New York City is the nation's largest transit market, with approximately one-third of all U.S. transit riders (Association 2015). Less well known is the city's large assortment of alternatives to subways, commuter rail and buses. Neighborhoods outside of the Manhattan core have long relied on informal networks of community cars, livery vehicles, commuter vans, dollar vans and other for-hire services. Each of these services tends to serve a particular niche, such as service between the city's three distinct Chinatowns in Manhattan and Queens (Tsai 2010). Yet formalizing these services has been difficult (King and Goldwyn 2014). While there are many reasons that formalizing informal transit is difficult, one factor is that these services are most used by immigrants and low income riders who always pay with cash.

Taxi services in New York City are regulated by the Taxi and Limousine Commission (TLC). In 2004 the TLC initiated a program that required all taxicabs to use technology that allowed for credit card processing, and also collected data about trip characteristics (King, Peters et al. 2012). This program was completed in 2008. In 2012 the city announced a program to increase the number of taxicabs outside of the Manhattan core into traditionally underserved neighborhoods. These new taxis, known as Green Cabs (because of their color) or boro cabs, cannot pick up passengers at the airports or Manhattan south of either 110th Street on the west side of Central Park or 96th Street on the east, and are available as either a street hail or pre-arranged ride. The full effect of the green cab program is not yet known for overall taxi access or ridership as the program is still new, but preliminary data can be used to assess how trips made in green cabs differ from those made in yellow cabs.[i]

Yellow taxicabs have been criticized for focusing their service on the airports and Manhattan's central business districts rather than serving the city as a whole. For years it was rare to see a taxi on the street of the outer boroughs (Brooklyn, Queens, Staten Island and the Bronx). This does not mean that these areas were not served. Rather, these areas were served by mix of informal and formal taxi services. Figure 2 is a photograph of a parking lot at a big box Target retail store in a large shopping complex in one of the unbanked neighborhoods in the Bronx. This is a typical scene at retail centers across the city, where a taxi queue forms to take people home once their shopping is complete. One green cab is available, but the balance of cars are licensed by the TLC and available for hire. In this shopping center the building management, the TLC and the licensed liveries worked together to create a queue in the parking structure to ensure orderliness. This arrangement is often provided in a language other than English, and nearly all of these passengers pay cash.

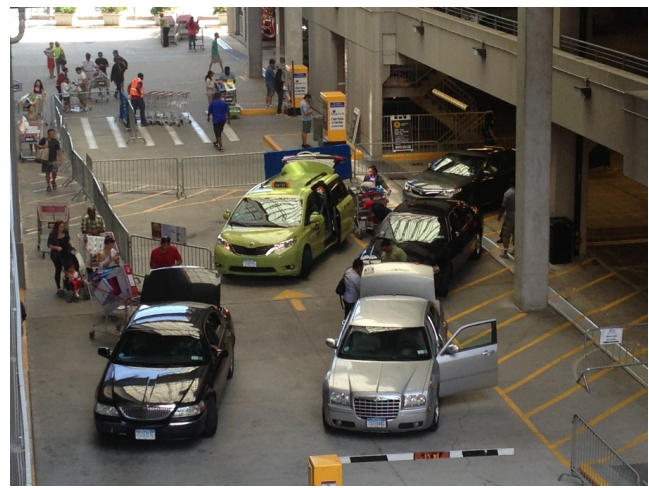

Figure 2: Community Cars at Shopping Center (Photo by author)

# 5. DATA AND METHODS

This research uses geolocated trip data from October 2014 for all yellow and green taxicabs in New York City, which were provided by request from the Taxi and Limousine Commission. The dataset is rich with trip information and includes trip origin, destination, time, number of passengers, fare paid, tolls paid, method of payment, tips (if paid by credit card) and other information. From the reported data distance traveled can be estimated but is not included in these analyses. These taxi data are combined with neighborhood level socio-economic data for analysis.

Table 2 shows the total trips by green and yellow taxis for the entire city during the study period. Yellow taxis make about ten times the number of paid trips as green taxis. This is for of many reasons, but primarily the yellow taxis are used much more intensively and there are simply thousands more of them. Each yellow taxi is used for two 12 hour shifts daily, and medallion owners are eager to keep drivers in the cabs to make sure they collect revenues. Green taxis, however, are typically owned by someone who drives part time and the leases the taxi for the balance of the week. The Green taxis are thus used for more flexible shifts.

The characteristics of trips by green and yellow can are quite different, as well. Obviously as green taxis cannot pick up in many areas of the city with high taxi demand overall trip characteristics are affected, but more importantly passengers use and pay for taxis differently. Fifty-five percent of all green taxis trips—serving outer boroughs by regulation—are cash fares. For yellow taxis, the likelihood of a cash fare is related to distance and whether the trip is an airport trip (these calculations are not shown).

| Table 2: Characteristics for All Trips, October 2014 | | |
|---|---|---|
|  | Green Taxi Trips | Yellow Taxi Trips |
| **Total Trips** | 1,491,266 | 14,232,488 |
| **Cash Trips** | 820,747 | 5,684,248 |
| **Share of Trips Paid Cash** | 55% | 40% |

Overall there are observable differences for cash payments by taxi type, location, trip origin and trip destination. It is impossible to know what characteristics differ between a typical yellow cab passenger and a typical green cab passenger, but something leads green cab passengers to use cash far more often than yellow cab passengers. The results shown on the maps (Figures 4-7) suggest that there is a spatial factor in play.

In Figures 4-7 the relative frequency of payment types by origin and destination for yellow and green taxicabs. In all maps there are stark lines that demarcate where riders predominately use cash (shown in yellow) and where they use credit (shown in blue). The areas marked with yellow are the places where cash is king. With the exception of a credit card hotspot surrounding Columbia University in Morningside Heights (a blue area circled in Figure 4) Manhattan payment types divide cleanly along income lines, where wealthy neighborhoods flanking Central Park (the empty white rectangle in the middle of the map surrounded by blue to the south and yellow to the north) on the Upper West Side and Upper East Side pay for taxi trips mostly with credit cards and poorer neighborhoods to the north in Spanish and Central Harlem are dominated by cash. One interesting aspect is that the socio-demographic characteristics of neighborhoods seemingly play a large role in determining payment type. It is likely that the cash or credit choice is a function of access to a bank account, for which these spatial data are a good proxy. Another takeaway is that much of the city still does not produce a lot of taxi trips and there is not enough data to present primary payment types at all.

Figure 4: Cash and Credit Payment Types for Green Cabs by Origin

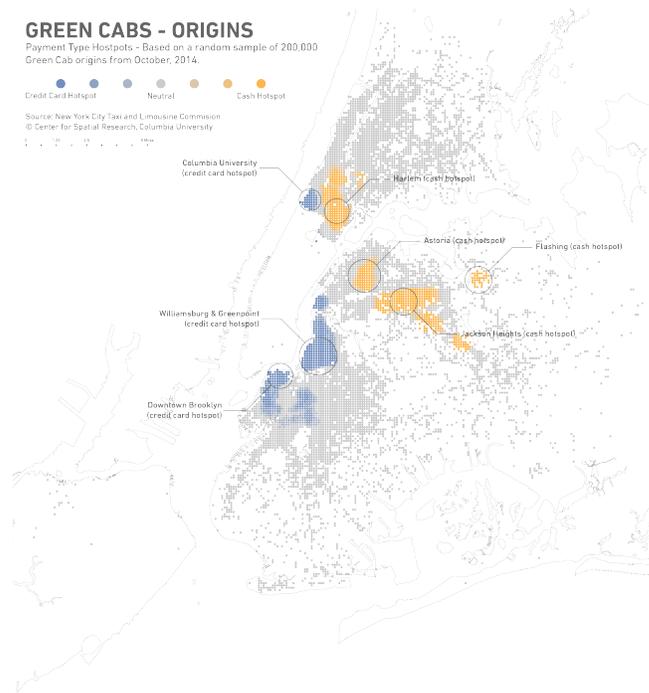

Figure 5: Cash and Credit Payment Types for Yellow Cabs by Origin

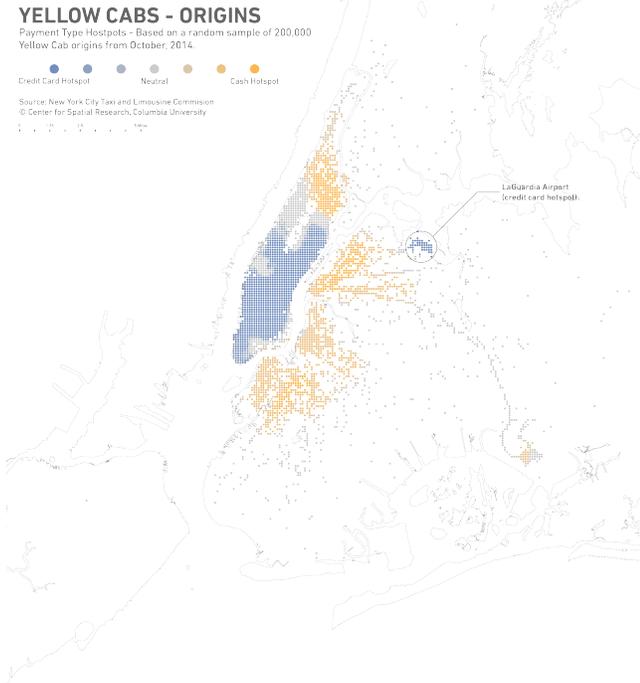

Figure 7: Cash and Credit Payment Types for Yellow Cabs by Destination

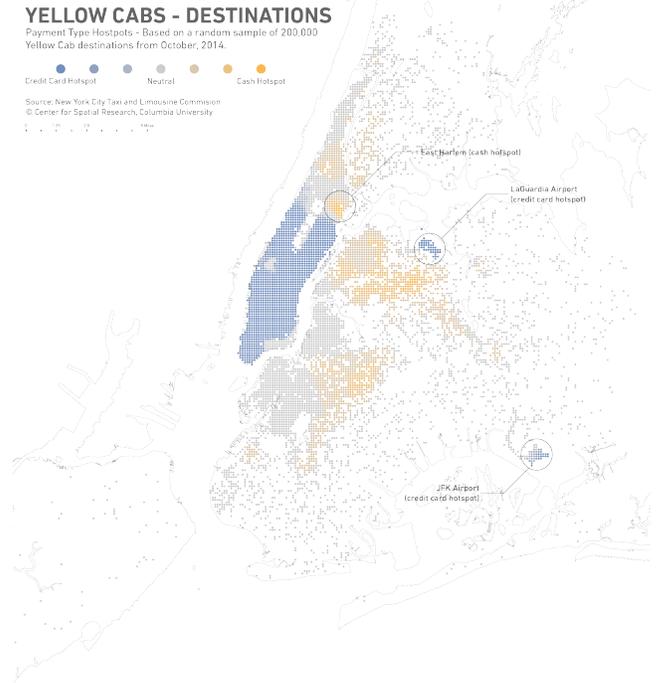

Figure 6: Cash and Credit Payment Types for Green Cabs by Destination

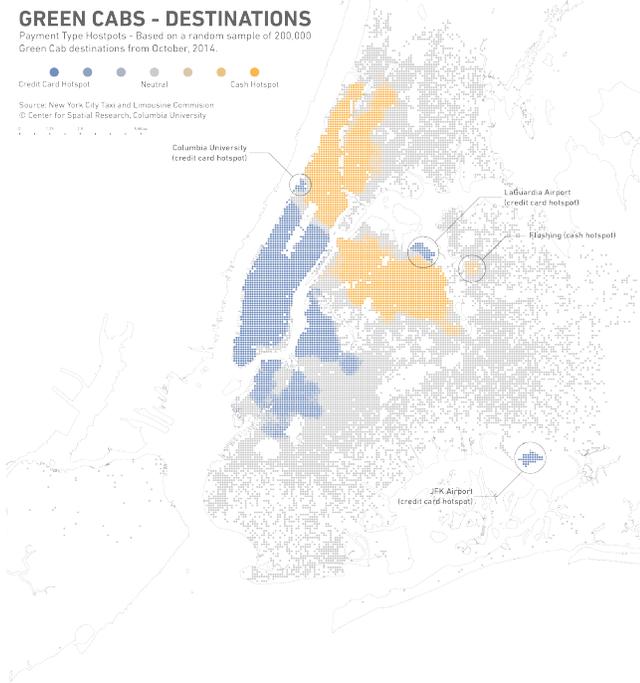

## 6. STATISTICAL ANALYSIS

In this section statistical analyses of the associations between socio-demographic characteristics and the share of cash fares for taxicabs are presented. The taxi trip data is limited in that it provides origins and destinations along with fare characteristics, but these data are associated with a known vehicle rather than a known passenger. As such, we can make a series of assumptions about a typical rider based on neighborhood factors. The regression analyses shown below used all trip data for the week of October 6-12, 2014, which is assumed to be a "typical week in terms of good weather, lack of holidays and no major school or employment breaks (n=3,217,092 for yellow taxi trips and n=330,024 for green taxi trips). These data are then assumed to represent a close approximation of the average trip, and thus the average trip taker. Taxi trips were aggregated spatially to the neighborhood level, which were the smallest geographies with available demographic data, and then analyzed by trip origins and destinations for cash payments. Origins and destinations were treated separately primarily because people leaving an area and returning to an area by taxicab may represent different groups of people.

Table 3 shows the summary statistics for dependent and independent variables considered for the regression models. These are shares of cash fares by origin (Ocash) and destination (Dcash) by neighborhood for all yellow and green taxi trips, and these are not mutually exclusive. Trips that begin and end in the same neighborhood will be counted as both Ocash and Dcash. For most of the outer borough neighborhoods the total number of intra-neighborhood trips is small as does not affect the overall results. For 2013, the percent of households in poverty, headed by a foreign born family member and unbanked are included.

The unemployment rate in 2013 was also considered but ultimately dropped from the analysis after post-test diagnostics.

**Table 3: Summary Statistics by Neighborhood (Percent)**

| Variable | Mean | Standard Deviation | Minimum | Maximum |
|---|---|---|---|---|
| Ocash | 0.42 | 0.07 | 0.35 | 0.81 |
| Dcash | 0.42 | 0.07 | 0.35 | 0.76 |
| Poor2013 | 0.18 | 0.10 | 0.03 | 0.40 |
| Foreignborn2013 | 0.38 | 0.12 | 0.17 | 0.64 |
| Unemployment2013 | 0.10 | 0.04 | 0.03 | 0.18 |
| Unbanked2013 | 0.13 | 0.08 | 0.03 | 0.31 |

Tables 5 and 6 show the regression results. The data are organized by neighborhood, and the dependent variable is either the share of cash trip by origin or cash trips by destination. Post-test diagnostics were used to evaluate multicollinearity, and the resulting models represent the best fit for the data. Ordinary least squares (OLS) is used along with generalized linear models (GLM), which accounts for the dependent variable not being normally distributed.

In all cases the strongest predictors of cash fares are the share of foreign born and the share of unbanked, and these effects are largest for taxi trip destinations. . These are large and positive coefficients that are highly statistically significant. The share of households in poverty is not statistically significant. In both OLS and GLM models the direction of effects and approximate magnitudes are similar, suggesting that both models adequately represent the relationships among variables. The r-sq for the OLS models suggest that close to half of the variation of cash fares by destination, which is a fairly high level of explanatory power for the model. It is likely that the reason poverty has an insignificant effect is that poverty is not a perfect predictor of banking status or immigrant status.

**Table 4: Regression Results for Cash Trips by Origin by Neighborhood**

|  | OLS | GLM |
|---|---|---|
| Poor 2013 | -0.431 (.339) | -1.814 (1.644) |
| Foreignborn2013 | 0.668 (.1333) | 2.808 (.461) |
| Unbanked2013 | 1.087 (.439) | 4.56 (2.139) |
| Constant | 0.2732 (.061) | -0.955 (.209) |
| F | 10.31 |  |
| r^2 | 0.39 |  |
| n | 52 | 52 |

**Table 5: Regression Results for Cash Trips by Destination by Neighborhood**

|  | OLS | GLM |
|---|---|---|
| Poor 2013 | -0.548 (.261) | -2.258 (1.220) |
| Foreignborn2013 | 0.587 (.102) | 2.415 (.359) |
| Unbanked2013 | 1.390 (.337) | 5.727 (1.602) |
| Constant | 0.241 (.047) | -1.065 (.144) |
| F | 17.52 |  |
| r^2 | 0.52 |  |
| n | 52 | 52 |

## 7. DISCUSSION

Taxicabs and for-hire transportation services are premium services that complement fixed-route transit and supply critical accessibility to people who do not or cannot drive. Ensuring that these services are available to all who need them is a desirable policy goal. What the data shown in this research shows is that in some cases access to bank accounts and credit cards may affect access to certain types of taxi services. There are strong correlations between neighborhoods with high shares of unbanked households and taxi trips—especially green cabs—paid with cash.

These results underscore an important aspect of emerging taxicab technologies, which is that many supporters of expanding taxicab supply base their support on the potential of new services to reach markets previously underserved. As potential can be refuted only through experience, existing firms in the taxi market look comparatively bad as the have a history that can be checked. It is a common claim that smart phone enabled taxi services will not employ the same geographic discrimination as conventional taxis because the drivers will respond to the service request. This is a fine idea, and a nice claim, and it may prove true at some point in the future. But many of the communities that need taxi services have high shares of unbanked households, who by definition cannot participate in a business that requires a credit card for access.

A scholarly example of this is a recent study by the BOTEC Analysis Corporation, where they sent researchers into various neighborhoods to check response times and total trip costs for taxicabs and Uber drivers (Smart, Rowe et al. 2015). The study is methodologically sound and the authors find quite conclusively that Uber cars arrive faster and cost quite a bit less on average. But in the Los Angeles neighborhoods not well served by taxis households have very high rates of being unbanked (Khashadourian and Tom 2007). These households also live in neighborhoods where carpooling acts as taxi services and is far more prevalent than taxis (Liu and Painter 2011), and Uber cars are likely slower and more expensive the taxi service actually used. It is possible that credit card based taxi services are simply out of reach for many of these communities.

Writers for the *Washington Post* collected data from Uber's API and found that Uber services offered faster service—measured by wait times after requests—to whiter and wealthier neighborhoods (Stark and Diakopoulos 2016). Such a claim is by itself not evidence of discrimination—and we want to be clear that is not part of our argument here—but taxicabs have long been subject to regulations in part to ensure access to service without regard to neighborhood, income, or race. While a systematic review of tech-enabled taxi services is beyond the scope of this paper, the studies cited above are suggestive that there may be spatial differences in taxi access even with app-enabled hailing.

The green taxicabs in New York City may have also helped solve one problem—taxi access—but introduced a new one—decline of community cars, which were shown in Figure 2. Community cars used to prowl the streets honking at prospective passengers, then the fare was negotiated for each trip. While this practice was illegal it was common. Through informal interviews with drivers and passengers of green cabs, some indicated they preferred the old system of negotiated fares—the green taxis have the same fare schedule as the yellow taxis—because drivers would give breaks to certain people, while other paid higher fares. Now the poorest riders, who previous could have negotiated a trip for whatever cash they were willing to pay, now have to pay the meter fare and it is often higher. As these are not data collected systematically through interviews the claims should be treated as speculation, but even as anecdotes they are insightful observations about how at least a few of the very poor riders made use of taxi-type services with cash.

One shortcoming of the taxi GPS data used is there is no specific information about the passenger. We can only assume that high rates of unbanked households is related to high rates of cash payments. While we feel this assumption is sound, the lack of passenger data limits the robustness of this and other analyses of taxi vehicle activities. We cannot say for certain a high share of unbanked households predict demand for cash payments for taxis, and this certainly requires additional surveys and passenger data. We also cannot evaluate these data for potential discrimination against passengers based on personal, locational or payment characteristics. There may be unobserved discrimination that affects the results shown.

With the green cabs in New York, it is not clear that unbanked people are underserved by taxicabs. However, this does not mean that taxi regulations and transportation policy shouldn't seek to protect vulnerable households. As the taxi industry goes through structural changes brought about by the rise of e-hailing applications, the city must consider ways to ensure access to all, not just those with a bank account.

## 8. CONCLUSION

This research presented an exploratory analysis of how taxi services in New York City exhibit market segmentation by fares payment methods. Overall, the green cabs, which were designed to serve outer boroughs and underserved areas, disproportionately have cash fares. The yellow and green taxi markets exhibit some aspects of market segmentation in that yellow cab trips in unbanked areas are more like yellow trips elsewhere and green cab trips are more like community cars and likely serve different riders. The use of cash to pay for taxi trips is strongly associated with neighborhoods that have high shares of unbanked and immigrant households. Airports and central business district taxi trips are more likely to use credit cards, and these riders likely have different socio-economic characteristics than outerborough riders. Some potential implications from these findings are discussed above, but the key points are worth reiterating. Discrimination in the taxi market is a long-standing concern. Taxi drivers are infamous for avoiding certain types of people and certain neighborhoods, which is a key argument in favor of public regulation against discrimination. Such discrimination should not be tolerated. A worry based on the analysis in this paper is that limiting taxi services to those with a credit card also leaves many households unserved, and may act as a new type of discrimination. Households on the edge of poverty go between having and not having bank accounts, and not having access to mainstream financial services may become a new type of discrimination without thoughtful policies.

## 9. REFERENCES

Association, A. P. T. (2015). APTA ridership report: third quarter 2015.

Doyle, J. J., J. A. Lopez and M. R. Saidenberg (1998). "How effective is lifeline banking in assisting the 'unbanked'?" Current Issues in Economics and Finance **4**(6).

Empowerment, N. Y. C. D. o. C. A. O. o. F. "Community Banking Profiles." from http://www1.nyc.gov/assets/dca/downloads/pdf/partners/Research-CFSS-BankingProfiles.pdf.

Glaeser, E. L., M. E. Kahn and J. Rappaport (2008). "Why do the poor live in cities? The role of public transportation." Journal of urban Economics **63**(1): 1-24.

Khashadourian, E. and S. Tom (2007). "The Unbanked Problem in Los Angeles." United Way of Greater Los Angeles.

King, D. A. and E. Goldwyn (2014). "Why do regulated jitney services often fail? Evidence from the New York City group ride vehicle project." Transport Policy **35**(0): 186-192.

King, D. A., J. R. Peters and M. Daus (2012). Taxicabs for Improved Urban Mobility: Are We Missing an Opportunity? Transportation Research Board 91st Annual Meeting, Washington, D.C.

Kneebone, E. (2010). The Great Recession and Poverty in Metropolitan America, Metropolitan Policy Program at Brookings.

Liu, C. Y. and G. Painter (2011). "Travel behavior among Latino immigrants: the role of ethnic concentration and ethnic employment." Journal of Planning Education and Research: 0739456X11422070.

New York City Department of Consumer Affairs (2008). Neighborhood Finanical Services Study.

New York City Department of Consumer Affairs (2013). Immigrant Financial Services Study.

Rhine, S. L. W. and W. H. Greene (2013). "Factors That Contribute to Becoming Unbanked." Journal of Consumer Affairs **47**(1): 27-45.

Sanchez, T. W. (2008). "Poverty, policy, and public transportation." Transportation Research Part A: Policy and Practice **42**(5): 833-841.

Sanchez, T. W., Q. Shen and Z.-R. Peng (2004). "Transit mobility, jobs access and low-income labour participation in US metropolitan areas." Urban Studies **41**(7): 1313-1331.

Sarlin, K. and E. Miller (2010). MORE THAN 825,000 ADULTS IN NEW YORK CITY DO NOT HAVE BANK OR CREDIT UNION ACCOUNTS ACCORDING TO NEW CITYWIDE STUDY. N. Y. C. D. o. C. Affairs.

Schuerman, M. (2015). Why MetroCards are Like Paper Towels. WNYC News.

Smart, R., B. Rowe, A. Hawken, M. Kleiman, N. Mladenobvic, P. Gehred and C. Manning (2015). Faster and Cheaper: How Ride-Sourcing Fills a Gap in Low-Income Los Angeles Neighborhoods.

Stark, Jennifer and Nicholar Diakopoulos (2016, March 10) "Uber seems to offer better service in areas with more white people. That raises some tough questions," Washington Post. Retreived from https://www.washingtonpost.com/news/wonk/wp/2016/03/10/uber-seems-to-offer-better-service-in-areas-with-more-white-people-that-raises-some-tough-questions/

Talukdar, D. (2008). "Cost of being poor: retail price and consumer price search differences across inner-city and suburban neighborhoods." Journal of Consumer Research **35**(3): 457-471.

Tsai, M. (2010). "The Chinatown Shuttle: Better than New York's Subway." Chinatown Stories Retrieved January 5, 2016, from http://www.chinatownstories.com/chinatown-shuttle/.

Yoh, A., H. Iseki, B. Taylor and D. King (2006). "Interoperable Transit Smart Card Systems: Are We Moving Too Slowly or Too Quickly?" Transportation Research Record: Journal of the Transportation Research Board **1986**: 69-77.

---

[i] The Green cabs were introduced at the same time Uber, Lyft and other competitors entered the market en masse. Since the growth of smart phone enabled services demand for Green cab medallions has declined and the city has not sold all available licenses.